# Origin and evolution of marginal basins of the NW Pacific：Diffuse-plate tectonic reconstructions

Junyuan Xu[a,*], Tom Kelty[b], Zvi Ben-Avraham[c], Ho-Shing Yu[d]

[a] *Department of Petroleum Geology, China University of Geosciences, Wuhan 430074, China.*
[*] *Corresponding author, E-mail address: jyxu@cug.edu.cn*
[b] *Department of Geological Sciences, California State University, Long Beach,* CA 90840, USA.
[c] *Department of Geophysics and Planetary Sciences, Tel Aviv University, Ramat Aviv 69978, Israel.*
[d] *Institute of Oceanography, National Taiwan University, Taibei, Taiwan.*

**Abstract**

Formation of the gigantic linked dextral pull-apart basin system in the NW Pacific is due to NNE- to ENE-ward motion of east Eurasia. This mainly was a response to the Indo-Asia collision which started about 50 Ma ago. The displacement of east Eurasia can be estimated using three aspects: (1) the magnitude of pull-apart of the dextral pull-apart basin system, (2) paleomagnetic data from eastern Eurasia and the region around the Arctic, and (3) the shortening deficits in the Large Tibetan Plateau. All the three aspects indicate that there was a large amount (about 1200 km) of northward motion of the South China block and compatible movements of other blocks in eastern Eurasia during the rifting period of the basin system. Such large motion of the eastern Eurasia region contradicts any traditional rigid plate tectonic reconstruction, but agrees with the more recent concepts of non-rigidity of both continental and oceanic lithosphere over geological times. Based on these estimates, the method developed for restoration of background diffuse deformation of the Eurasian plate and the region around the Arctic, and the related kinematics of the marginal basins, we present plate tectonic reconstruction of these marginal basins in global plate tectonic settings at the four key times: 50, 35, 15 and 5 Ma. The plate tectonic reconstruction shows that the first-order rift stage and post-rift stage of the marginal basins are correlated with the first-order slow uplift stage and the rapid uplift stage of the Tibetan Plateau, respectively. The proto-Philippine Sea basin was trapped as a sinistral transpressional pop-up structure at a position that was 20°south of its present position. While the Japan arc migrated eastward during the rifting period of the Japan Sea basin, the Shikoku Basin opened and the Parece Vela Basin widened.

## 1. Introduction

Development of a gigantic dextral pull-apart rift system along the NW Pacific margin indicates that there was a large amount of northward movement of the South China block and compatible movements of other blocks in eastern Eurasia and region around the Arctic during its rifting period. The northward plate movement was predominantly the product of the collision between the Indian and Eurasian plates during the rifting period of the basin system (Fig. 1) (Xu et al., 2012). However，such a large displacement contradicts any traditional rigid-plate tectonic reconstruction. This paper estimates the displacement amount and presents plate tectonic reconstructions of the basin system together with the Philippine Sea basin that challenges traditional views on the history of movement of the Eurasian and North American plates.

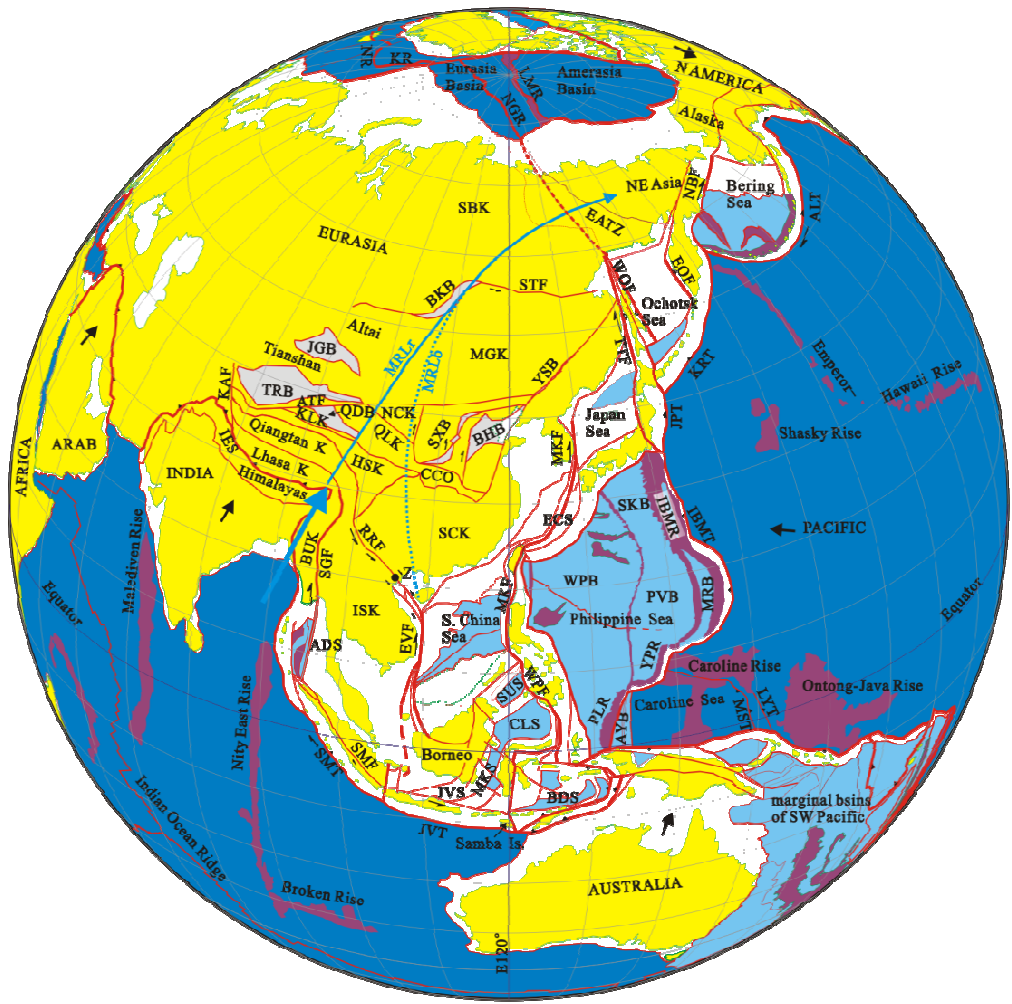

Fig. 1. Tectonic sketch of the marginal basins of NW Pacific, tectonic division of the Large Tibetan Plateau (LTP) and general motion circuits of the east Eurasian plate and region around Arctic (EUAR) during the dextral pull-apart rifting of the marginal basins. Yellow, white, ice blue, sky blue and purple areas denote land, shelf-slope (or transitional crust), marginal basins' oceanic crust, open oceanic crust and rises in oceanic crust, respectively. Dotted and solid cyan curve lines with an arrow are the model background and possibly real maximum rotation lines (MRLb and MRLr), respectively. Straight thick arrow denotes the largest motion vector of India. About zero strike-slip displacement on the RRF is at the point Z. Name abbreviations (alphabetically): (1) basin (-B) or sea (-S): ADS=Andaman, AYB=Ayu, BDS=Banda, BHB=Bohai Gulf, BKB=Baikal, CLS=Celebes, ECS=East China, JGB=Junggar, JVS=Java, MRB=Mariana, MKS=Makassar, PVB=Parece Vela, QDB=Qaidam, SKB=Shikoku, SUS=Sulu, SXB=Shanxi, TRB=Tarim; WPB=West Philippine, YSB=Yishu; (2) block (-K, or K): BUK=Burma, HSK=Hoh xil - Sangpan-Ganze, ISK=Indochina-Sumatra, KLK=Kunlun, MGK=Mongolia; NCK= North China, QLK=Qilian, SCK=South China;  (3) fault (-F): ATF=Altyn Tagh, EOF=East Okhotsk, EVF=East Vietnam, KAF=Karakoram, MKF=Manila-E. Korea, NBF=NW Bering, SGF=Sagaing, RRF=Red River, SMF=Sumatra, TTF=Tartar-Tanakura, WOF=W. Okhotsk, WPF=W. Philippine; (4) ridge or rise (-R): CAR=Caroline, IBMR=Izu-Bonin-Mariana, KR=Konipovich Ridge, KPR=Kyushu-Palau, LMR=Lomonosov Rise, NGR=Narsen-Garkel Ridge, PLR=Palau, YPR=Yapu; (5) trench (-T): ALT=Aleutian, IBMT=Izu-Bonin-Mariana, JPT=Japan, JVT=Java, KRT=Kuril, MST=Mussau, SMT=Sumatra; (6) others: CCO=Central China orogen, EATZ = the transitional zone between the Eurasian and North American plates.

2. **Estimation of movement of eastern Eurasia during the Cenozoic**

**2.1. Amount of extension of the marginal basin system**

As shown in Fig. 10 (b) of Xu et al. (2012), the South China block could move northward by 10.04° to 11.25° relative to Samba Island of East Java Arc, which was enough to produce the dextral transtension of the South China Sea basin and the Java-Makassar-Celebes-Sulu Seas (JMCS) basin system except most of opening of the Celebes Sea. If Samba Island is fixed to the hotspot reference frame or paleomagnetic reference frame for absolute plate motions (e.g., Muller et al., 1993; Torsvik et al., 2008), the amount of extension when the pull-apart basin system formed is equal to absolute northward motion of the South China block during the rifting period.

However, the absolute motion of the block could have been larger or even some less than if Samba Island moved northward relative to these reference frames. It should be pointed out that the lithosphere is not completely plastic but elastic-plastic, the southeastern part of the Indochina-Sumatra block and the east Java arc together with West Sulawesi could move southward in a small scale when rotated CW, as shown in non-rigid analog modeling (e.g., Fournier et al., 2004). In other words, the Java Trench could retreat southward.

In order to better understand the displacement amount and history of the east Eurasia, the other evidence for these movements should be considered.

**2.2. Paleomagnetic evidence**

As is well-known, paleomagnetic data can be used to reconstruct the movement of plates or blocks. However, paleomagnetic data sets often contradict each other and have large errors and multiple solutions that are caused by many factors (e.g., local re-magnetization, shallowing of inclinations by overburden or steepening of inclinations by horizontal tectonic compression, abnormal declinations by local deformation rotation, non-dipolar geomagnetic field and even true polar wandering). Movements of blocks or plates that are predicted using these data should therefore be constrained with geological data. Generally paleomagnetic data support or don't contradict the large movements of east Asia since 50 Ma which is presumed to be the initial time of the collision of India with Asia (see next section for detail).

The paleomagnetic data from the South China block suggest that it moved northward 10° to 12° since 50 (or 40) Ma and then shifted southward or remained stable (McElhinny and McFadden, 2000；Clyde et al., 2003), but earlier authors report that the northward latitude shifts were 8° between 50 to 15 Ma and then -4° (*negative means "southward", which will be same in following text*) for the eastern South China block (Liu et al., 1990; similar results see Yuan et al., 1992) and 19.4°±6.3° since Paleocene to Eocene for the west South China block (Huang and Opdyke, 1992) (see summary of Cogne, et al., 1999) .

Jin et al. (2004) inverted the magnetic anomalies of the seamounts in the South China Sea and showed that the nine seamounts in the Central sub-basin moved northward 3.09°, 2.99°, 5.04°, 3.53°, 7.79°, 9.48°, 9.94°, 11.06° and 7.77°, respectively, from a spreading center (located at about N14.5°) progressively to about N17.7°, and the seven, randomly located seamounts in the SW sub-basin moved southward 4.52° on average. The exact ages of the seamounts are uncertain, but clearly show the northern oceanic crust had large-scale, northward movement and probably later moved southward.

Scant and contradictory Cenozoic paleomagnetic data from the Indochina-Sumatra block cannot fully display its Cenozoic movements. According to paleomagnetic data from Mae Mho Basin (18.32°N, 99.74°E), Phetchaburi Basin (13.16°N, 99.67°E) and Krabi Basin (8°N, 99.05°E) in the west of the Indochina-Sumatra block (Richer *et al.*, 1993), the three basins moved northward 11.84±9.91°, 8° and -7.5° respectively, since Oligocene-Miocene. The -7.5° contradicts the basic tectonic history of the block. The inclinations from this block, especially its NW part close to the Red River fault zone, were probably made steeper by transpressional deformation.

Fujita et al. (1987) summarized that southwest Japan drifted 4° ($\alpha_{95}$ not given) north since the early Paleogene, and Tosha and Hamano (1988) reported Oga Island (40°N, 140°E) of NE Japan moved northward about 8.4±10° since about 52 Ma. According to paleomagnetic inclinations reported by Otofuji et al. (1995, 2002) from the Shihote-Alin region of the Mongolia block, north of the Japan Sea, moved northward about 6.7 to 14° since about 50 Ma.

Gilder *et al.* (1996) proposed that there is the possibility of northward movement of the Siberia block relative to Europe along the Ural orogenic belt due to the India-Eurasia collision on basis of paleomagnetic and geological data. Cogne et al. (1999) summarized that the Lhasa block (30°N 91°E), the Qiangtang block (32.8°N 96.6°E), the Tarim block (43°N 90.5°E or 37.7°N 79°E) and the Junggar block (44.2°N 86°E) south of the Altai orogen drifted northward by 16.6°±6.4°, 15.3°±6°, 22.2°±16.7° (or 10.3°±12.7°) and 12.2°±6.4°, respectively, since the Paleocene to Eocene. Cogne et al. (1999) further concluded that predictions of positions of the Siberia block based on the Apparent Pole Wonder Path (APWP) of Eurasia (e.g., Besse and Courtilott, 1991) is erroneous, and suggested that Eurasian plate might not be rigid and could be divided into three sub-plates that are located between the Urals and Tomquist-Tesseyre Line, respectively. The possible northward drift of 12.2°±6.4° of the Junggar block manifests that the southeast Siberia block and the western Mongolia block must have moved northward because the Altay orogen between the southeast Siberia and Junggar blocks could not accommodate large shortening of 12.2°±6.4°. This assumes that the western Siberia block had remained stable. In their Cretaceous palaeographic reconstruction, there is not a mega-fault between the Junggar-Mongolia region and the Siberia block in the Cenozoic as hypothesized by Halim *et al.* (1998). If the mega-fault had existed, the region of the Okhotsk Sea would have been intensively compressed. Besides, the great northward movement of NW China, north China would also have to have moved northward. However, we have found no evidence to suggest that a major dextral fault (or ductile shear zone) existed between northwest China and north China.

Hankard et al. (2007) report that Paleogene and Neogene paleomagnetic data from Siberia and Mongolia indicated that their paleomagnetic poles at 13 and 20 Ma (16.5 Ma on average?) are fairly consistent with those of the reference APWP for Eurasia (Besse and Courtillot, 2002). However, the 30 Ma pole appears far-sided with respect to the corresponding reference pole. They suggest that Siberia was located 1000 km south of the predicted position at 30 Ma and the Eurasian plate underwent non-rigid deformation.

The paleomagnetic data from White Pass in the northern North American Cordillera indicate a paleolatitude discordance of 8±4° south and a CW rotation of 40±9° with respect to the North American craton since 50 Ma (Symons et al., 2000). The northernmost North American Cordillera possibly moved southwards relative to the North American craton, responding to the eastward movements of the region around the Arctic.

Recently, Dupont-Nivet et al. (2010) provide new paleomagnetic results from volcanic rocks from Mongolia, and sedimentary data sets from China corrected for inclination shallowing, which together with compiled reliable Asian data sets confirm that Asian paleolatitudes are 5–10° lower than predicted by the APWP in the 50–20 Ma period. Dupont-Nivet et al. (2010) investigated two explanations: (1) Asian was indeed >1000 km further south than predicted by the APWP (due to Eurasian non-rigidity, inaccurate plate circuit for Eurasia, or inaccurate global APWP) or (2) large and long-standing time dependent octupolar contributions (up to 16%) to the geomagnetic field. We favor the first explanation.

In conclusion, the above paleomagnetic data show that the South China block possibly moved northward by about 10° and the other blocks of the eastern Eurasia and the region around the Arctic had compatible motions since 50 Ma.

## 2.3. Shortening history and deficits in Tibetan Plateau

### 2.3.1. Time of the initial collision

Age of the initial collision between India and Eurasia has been contentious for the past decades and its estimates range from the Late Cretaceous (> 65 Ma) to the Late Eocene (<40 Ma) (see summary of Rowley (1996) and Henderson et al. (2011)). Most authors believe that the collision started about 55 to 50 Ma ago (e.g., Powell and Conaghan, 1973; Patriat and Achache, 1984; Hall, 2002; Leech et al., 2005; Royden et al., 2008; Chen et al., 2010). Some authors (e.g., Aitchison et al., 2007) reject "the around 55 Ma dogma of the initial collision time" and propose the continent-continent collision is very young (about 34 Ma ago). However, with regard to Aitchison et al. (2007), Garzanti (2008) concludes that among the numerous unsolved chronic problems that afflict Himalayan geology, the date of arrival of the Indian continental margin at the Transhimalayan trench at 55 Ma stands as one of the few major events that are robustly constrained by multidisciplinary geological evidence. Najman et al. (2010) made a detailed discussion on ideas of Aitchison et al. (2007) and come to about the same conclusion as Garzanti (2008), suggesting the initial time is about 50-52.8 Ma along the central segment of the India-Asia suture. Therefore, 55 to 50 Ma seems more convincing as a time of the initial collision.

However, the exact time of the initial collision and time of the "*full*" collision between India and Asia are very difficult to determine. The motion rates of India-Eurasia "convergence" rate (hypothesized that Eurasia was rigid and fixed) and rates of absolute motion of India (e.g., relative to Fixed Hotspot Frame) vary with authors (e.g., Torsvik et al., 2008; Molnar and Stock, 2009, Copley et al., 2010). The time of the first drop in these rates could be 55, 50, or even 47.1 Ma (Fig. 2). The rate change alone cannot determine the exact initial collision time. The rate change of a plate not only depends on its interaction with other plates, but also on the state of convection of asthenosphere below related plates or even global plates and other factors (e.g., van Hinsbergen et al., 2011; White and Lister, 2012). For example, India moved in very different rates from its rifting away from Gonwana to its collision with Asia (e.g., Fig. 2(a)), but no other plate helped or resisted its motion during this period of time. Cessation of marine facies deposition, located between India and Asia, could be a good approximate indication of the collision time (e.g., Rowley, 1996; Najman et al., 2010). The kinematical balance of matter between India and Asia could be another good indicator of the collision time (see 2.3.3 for further discussion). Taking the time of rate change in the India's motion, the cessation of the marine facies deposition and the kinematic balance of matter between India and Asia, we agree with the diachronous collision along the suture and propose that India touched with Asia along the central-eastern segment of the suture at about 50 Ma and the collision propagated westward with the full collision beginning at about 45 Ma (see Section 3 for further detail).

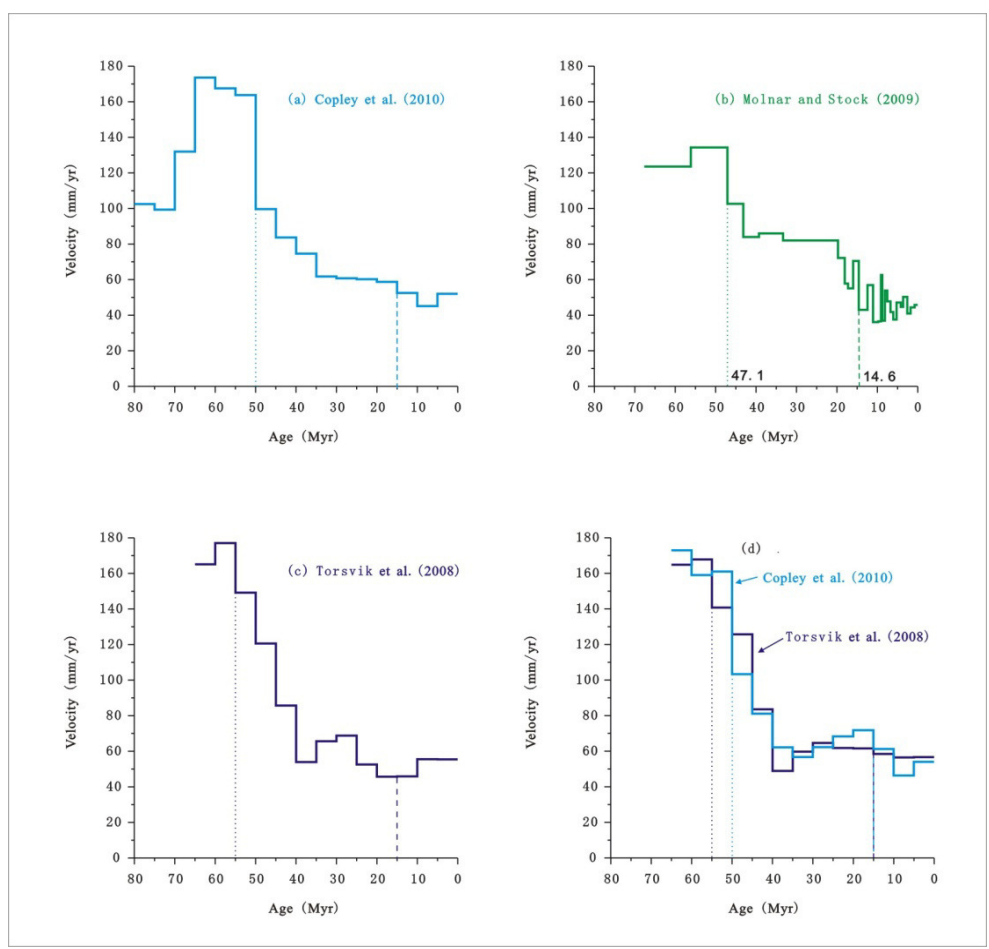

Fig. 2. (a) - (c) India-Asia convergence rates at the reference point (28°N, 90°E) calculated on basis of data provided by Copley et al. (2010), Molnar and Stock (2009) and Torsvik et al. (2008), hypothesizing that Eurasia plate is rigid and fixed. (d) Rates of India relative to the African Hotspot Fixed Frame according to data of Torsvik et al. (2008) and Copley et al. (2010). Dotted lines denote time of the first sharp drop of these rates, and dashed lines denote change in theses rates around 15 Ma.

### 2.3.2. Shortening history

The Himalayas and the Tibetan Plateau and the high-altitude region (Tarim-Tianshan-Altai) north of the plateau (taken together) may be designated as the Large Tibetan Plateau (LTP) (Fig.

1). Shortening in the LTP has continued since the initial collision between India and Asia. Uplift of the Tibetan Plateau was formed by shortening and thickening of its crust (e.g., Dewey et al., 1988). Although the uplift of the plateau could have resulted from thermal convection or delamination of upper mantle (e.g., Molnar et al., 1993; Chung et al, 1998), the recent studies (e.g., McKenzie and Priestley, 2010) find that cold and thick (about 300 km) lithosphere is present everywhere beneath the Tibetan plateau, and deny the thermal-related uplift mechanism. The uplift is simply related to thickening of the crust or lithosphere and the shortening history of the LTP can be shown by its uplift history, if erosion is not considered.

However, the uplift history of the LTP has been very controversial and many different hypotheses have been proposed (e.g., Powell and Conaghan, 1973; Patrait and Achache, 1984; Mercier et al., 1987; Dewey et al, 1988; Molnar and England, 1990; Wang and Coward, 1990; Harrison et al., 1992; Coleman and Hodges, 1995; Li, 1995; Fielding, 1996; Le Pichon et al., 1997; Murphy et al., 1997; Yin and Harrison, 2000; Tapponnier et al., 2001; Spicer et al., 2003; Rowley and Currie, 2006; Zhou et al., 2006; Wang et al., 2007). It is difficult to find which of these hypotheses is most accurate. But there are some convincing facts as follows.

(1) There were episodic uplifts in the LTP from the initial collision time throughout to the present. The episodic stages can be divided according to unconformities or sedimentary responses in the Cenozoic basins.

(2) The southern, central and northern Tibetan Plateau achieved its elevations close to today or underwent a significant uplift before the Eocene-Oligocene transition (EOT) *(*about 35 Ma, maximum 38 Ma, minimum 32 Ma) (e.g., Yin and Harrison, 2002; Dai et al., 2005; Rowley and Currie, 2006; Dupont-Nivet et al., 2008; Wang et al., 2007; Yin, 2009). This means that there existed either more rapid uplift between time of the initial collision (55 Ma) and 35 Ma if its pre-collision elevation was low (e.g., the elevation is zero or less than 500 m) (e.g., Dewey et al, 1988), or slow uplift if its high elevation was created before 55 Ma (e.g., Murphy, 1997; Kapp et al., 2005).

(3) The Tibetan Plateau had a relatively slow deformation episode either between about 35 to 15 Ma or between 55 to 15 Ma. It is worth mentioning that around 25 Ma there was a tectonic event that caused a wide-distributed unconformity between Paleogene and Neogene both in the Tibetan Plateau and in the marginal basins of the NW Pacific. This short tectonic event possibly did not cause evident uplift in the LTP. There could be some uplift in the northern Tibetan Plateau (around the Qaidam and Tarim basins) between 35 to 15 Ma, but the shortening rate was quite lower and the rapid uplift occurred after 15 (or 12) Ma (e.g., Zhou et al, 2006). There was more intensive tectono-magmatic activity in Himalayas between 35 Ma and 15 Ma (e.g., Dewey et al., 1988; Yin and Harrison, 2000).

(4) Intensive tectonic activity in the LTP, rejuvenated or initiated since 15 Ma (e.g., summary of Molnar and Stock (2009)). It is here emphasized that compression between the eastern Tibetan Plateau and the Sichuan basin initiated (e.g., Kirby et al, 2002; Wang and Meng, 2009) and most of NS-trending grabens in the Tibetan Plateau formed after about 15 Ma (e.g., Blisniuk, 2001; Kapp and Guynn, 2004) as well as the Karakoram fault activated since about 14 Ma (e.g., Bhutani et al., 2003).

Based on the above basic facts related to uplift of the LTP and the rifting history of the marginal basin system of the NW Pacific, we suggest two first-order stages for uplift of the LTP:

(1) A slow uplift stage. This stage lasted from around 55 to 15 Ma when intensive rifting of

the marginal basin system took place. During this stage, both the India-Asia convergence and the Tibetan Plateau uplift were generally slow. But the uplift and convergence could have been more rapid from 55 to 35 Ma than from 35 to 15 Ma.

(2) A rapid uplift stage. This stage lasted from 15 Ma to the present when intense rifting of the major marginal basins generally ceased. During this stage, the rapid uplift of the plateau was simply due to the rapid convergence. Uplift and thickening of the Tibetan crust could have occurred in one of ways as suggested by some authors (e.g., Powell, 1986; England and Searle, 1986; Zhang and Morgan, 1986) or in a combination of these ways, which depends on the uplift stages and the places, and will be discussed in the section 3.3.

### 2.3.3. Shortening deficits

Recent studies have found an important but puzzling phenomenon related to the shortening of the LTP: A large shortening deficit in the LTP, that is, the shortening amount of the LTP is much less than the India-Asia convergence that is predicted based on the plate reconstructions (e.g., Le Pichon et al., 1992; Johnson, 2002). We suggest the shortening deficits mainly result from the large northward motion of the eastern Eurasia and can be used as an independent kinematical parameter to make new plate tectonic reconstruction of the marginal basins of the NW Pacific.

The shortening amount budgets of the LTP, depending on its paleo-elevation (or thickness of its crust or lithosphere) at the initial collision, vary with authors (e.g., Dewey et al., 1988; Le Pichon et al., 1992; Yin and Harrison, 2000; Xu and Zhang, 2000; Johnson, 2002). Among the previous works, Le Pichon et al. (1992) most systematically estimated the shortening amounts and shortening deficits in a straight-forward way. Their basic dataset combined with *the shortening history of the LTP (stated above)* are used by us to further estimate the shortening amount. On the other hand, we use the rotational parameters of several recent Indian and Eurasian plate reconstructions (Torsvik et al., 2008; Molnar and Stock, 2009; Copley et al., 2010) together with the initial collision time (stated above) to predict the convergence between India and Asia.

If the initial collision time is 50 Ma (stated above) and the LTP's average elevation at 50 Ma is zero meters along every line, the shortening deficits along all the four lines are more than 1330 km, with the maximum reaching 1980 km (Fig. 3). If the LTP's average elevation is 500 m at 50 Ma, the deficits are more than 1820 km, reaching a maximum of 2410 km (Fig. 4). More than 500 m average elevation at 50 Ma will result in unrealistic shortening deficits. The real LTP's average elevation should be between 0 m and 500 m (but the different blocks could have different elevations). According to Le Pichon et al. (1992) (Fig. 2 and Fig. 3, or Fig. 1(inset, right), if the initial collision time is 45 Ma the deficits are more than 700 km and 1200 km for 0 m and 500 m topography base levels respectively. Only if the initial collision time is 35 Ma and the preexisting topography level is zero, the deficits could be neglected according to Torsvik et al. (2008)'s reconstruction. However, such young initial collision time of 35 Ma has been denied by the reasoning stated above. It should be noted that, even if the initial collision time was really 35 Ma, but if elevation of the Tibetan Plateau before 35 Ma was the about the same as its present elevation (e.g., Wang et al., 2007), the shortening deficits in the LTP would also be very large (1000's km). If the initial collision time is 65 Ma, that large shortening deficits seem unrealistic as well.

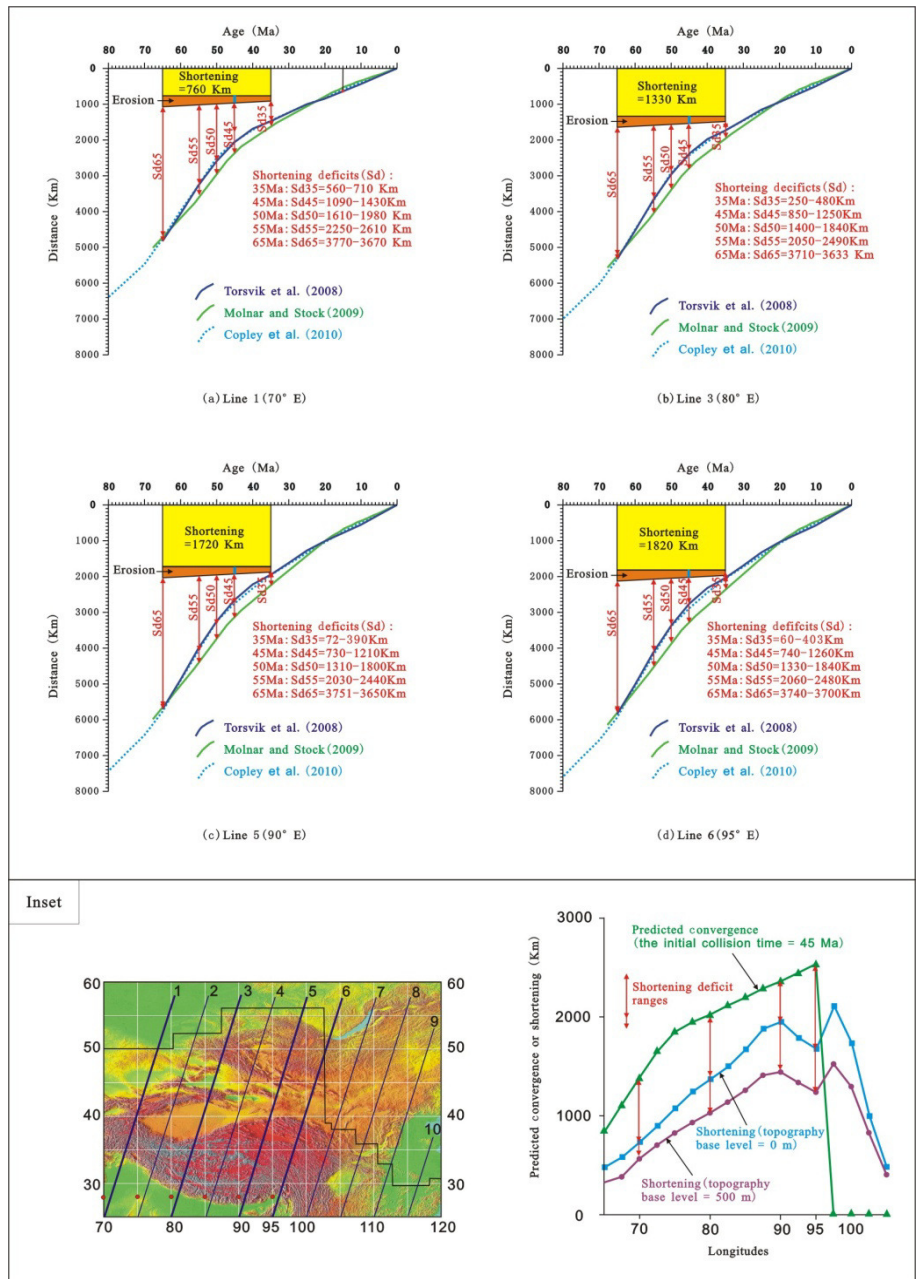

Fig. 3. Shortening deficit estimates along the four lines, Line 1, Line 3, Line 5 and Line 6 (see the lower inset (left) for their locations), with the topography base level hypothesized to be zero meters. Yellow areas show the shortenings that are calculated based on the present and the topography base level according to the method proposed by Le Pichon et al. (1992), within the possible initial collision time widow ranging from 35 to 65 Ma. Orange areas show erosion corrections, which equal to 200 km at 45 Ma (blue bars. Le Pichon et al. (1992)) and linearly change with time. The predicted convergences or shortenings between India and Eurasia at given reference points (red dots at about the southern ends of the lines, respectively) versus time are calculated according to Torsvik et al. (2008), Molnar and Stock (2009) and Copley et al. (2010). The red arrows show shortening deficit ranges at the five given initial collision times (65, 55, 50, 45 and 35 Ma) on basis of Torsvik et al. (2008) and Molnar and Stock (2009) (Copley et al. (2010)'s are very close to Torsvik et al. (2008)'s). The lower inset (left) and (right) show the locations of the four lines among ten lines, and the predicted convergence and the real shortenings along the lines (after Le Pichon et al. (1992). the colored topography is after NASA. See Le Pichon et al. (1992) for detail), respectively.

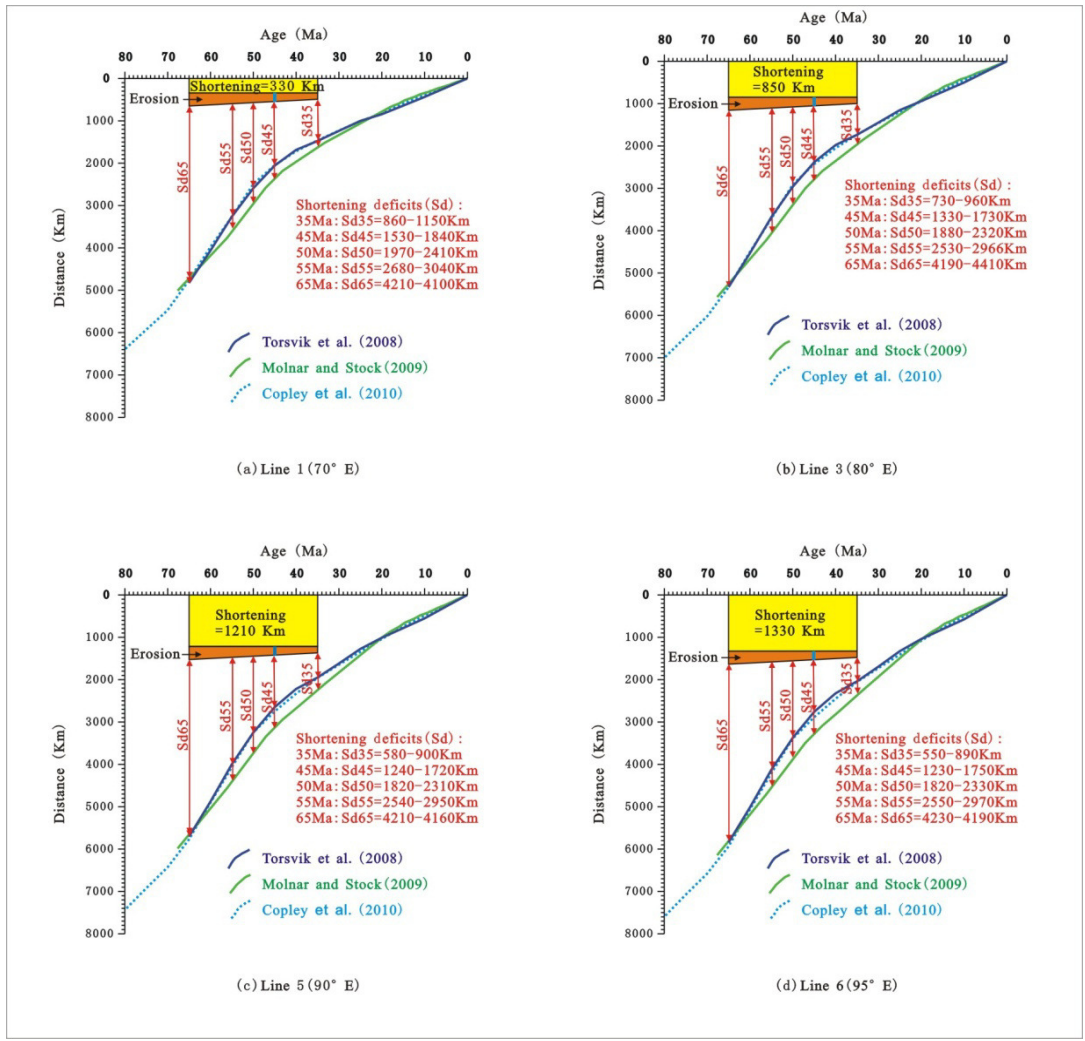

Fig. 4. Shortening deficit estimates along the four lines, Line 1, Line 3, Line 5 and Line 6 (see the lower inset (left) of Fig. 1 for their locations), with the topography base level hypothesized to be 500 m. All the legends are the same as in Fig. 1.

Where is the missing crust? Le Pichon et al. (1992) propose that a combination of transfer of the lower crust to the mantle through eclogitization and the lateral extrusion could account for a minimum of one third and maximum of one half of the total amount of the shortening between India and Asia (since 45 Ma). They also believe that the excess mass west of the eastern syntaxis that resulted from the lateral extrusion does not account for one third to one half of the deficits west of the eastern syntaxis. Some other authors also postulate that the deficits mainly result from the lateral extrusion, supporting the collision-extrusion tectonics proposed by Tapponnier et al. (1982) (e.g., Jaeger et al., 1989; Tapponnier et al., 2001; Royden et al, 2008). However, Dewey et al. (1988) and England and Molnar (1990) and others (e.g., Clift et al., 2008; Hall, 2008) reject any great magnitude of lateral extrusion after thorough analysis of tectonic history of the region around the Tibetan Plateau. Dewey et al. (1988) estimated that, if any, the lateral extrusion is less than 15% of total east-west width of the Tibetan Plateau. After re-examination of the recently published data from the plateau, we support Dewey et al. (1988) and England and Molnar (1990), emphasizing that the uplift and crustal thickening of the region east of the eastern syntaxis was mainly caused by the northward motion of the western Indochina-Sumatra block rather than the lateral extrusion.

Eclogitization could account for a part of the shortening deficits. Le Pichon et al. (1992) conclude that if a significant part of the lower crust has been eclogitized, the amount of "lateral extrusion" could be reduced to as little as 10%, the minimum amount compatible with the eastward transfer of Tibetan crust. But in contrast Le Pichon et al. (1997) proposed the eclogite in the lower crust changed into granulites and caused the Neogene uplift of the Tibetan Plateau. Although some part (e.g., 30%) of the lower crust in the LTP may have been eclogitized (e.g., Schulte-Pelkum et al., 2005; Hetényi et al., 2007), it can only account for a very small part of such large deficits as shown in Fig. 3 and Fig. 4 if the initial collision time is 50 Ma or earlier.

It is noted that if the time of the initial collision along the whole suture is same (50 Ma), the shortening deficits in the westernmost part of the LTP (e.g., Fig. 3(a) and Fig. 4(a)) are a little larger than in the easternmost part (e.g., Fig. 3(d) and Fig. 4(d)), but both the real and predicted shortenings in the westernmost part are much smaller. It is more reasonable that more predicted convergence has more shortening deficits. The initial collision may be diachronous and could be 45 Ma in the westernmost part.

It is conclusive that at least 1200 km of the shortening deficits in the easternmost part and about 1000 km in the westernmost part are robust although exact shortening deficits cannot be known at the present knowledge's level. The shortening deficits are simply attributed to the NNE-NE-ENE-EW movement of east Eurasia and region around the Arctic in step progressively approaching North America. The "missing" crust indicated by the shortening deficits extrudes into the NW Pacific.

## 3. Plate tectonic reconstructions of the marginal basins

### 3.1. Non-rigid deformation of plates

Geological and geophysical data indicate deformation of both continental and oceanic plates is non-rigid in some cases and some diffuse plate boundaries in both continents and oceans exceed a length of 1000 km on a side (e.g., Dewey and Burke, 1973; Ratschbacher and Ben-Avraham, 1995; England and Molnar, 1997; Gordon et al., 1998; Kronenberg et al., 2002; Hall, 2002).

Several lines of evidence suggest that the lithosphere of the continents may act like a viscous fluid over geological timescales just as the mantle does (e.g., Oxford University, see **http://www.earth.ox.ac.uk/~geodesy/research.html**). Most of all the numerical modelling of deformation of the LTP is made under the non-rigidity of continental plate (e.g., England and Houseman, 1986; Royden et al., 1997; Yang and Liu, 2009). In addition, the great difference between the paleomagnetic latitudes from east Asia and the paleolatitudes expected based on APWP of the Eurasian plate, as mentioned above, also support its non-rigid deformation during the Cenozoic.

Non-rigid deformation of the oceanic crust of some marginal basins was advocated by several workers (e.g., Mrozowski and Hayes, 1979; Gordon, 1998). We postulate that transform faults within oceanic plates can accommodate mega-scale, non-rigid shearing deformation whose principal shear can be either sub-parallel to the faults or sub-perpendicular to the faults.

Mechanically, the non-rigid deformation of the upper crust of a plate occurs easier when it is horizontally compressed or sheared than when it is stretched (or spread). If a plate is compared to an ice sheet, it is rigid like flowing ice sheet when driven by underlying flowing water (like mantle convection), and it may be non-rigid when driven by gravity (some like compression or collision on the lateral sides). Non-rigid plate deformation in geological history is possibly much more common than previously recognized. It should be noted that diffuse horizontal nearly-simple shears over a wide range of region are not easy to identify because these shears are close to plane strain deformation and can cause minor uplifts or subsidence in topography. Also, the non-rigid deformation may be accommodated by strike-slip faults and fractures.

It is emphasized that the present "rigidity" of some plates or blocks (e.g., the Siberia Block) indicated by GPS and similar data (e.g., Smith et al., 1990; Larson et al., 1997; Heflin et al., 2004) does not necessarily mean their past rigidity. The present motions of plates as shown by the NUVEL-1A and MORVEL models (e.g., DeMets et al., 1994, 2010) generally represent those of the past 3-6 Ma and perhaps even the past 15 Ma, when the marginal basins of NW Pacific basically stopped spreading. Nevertheless, GPS and similar data show that there are not only significant differences between the models and the measurements at most plate boundaries, but also in some cases at considerable distances from the boundary (e.g., Smith and Baltuck, 1993).

Understanding continental and oceanic plates' non-rigidity is of key importance to understanding global tectonic settings of formation of the gigantic linked dextral pull-apart rift system of the NW Pacific.

## 3.2. Method of reconstruction of non-rigid deformation of the Eurasian and North American plates

Reconstruction of rigid plate motion on the sphere is made using the Euler Theorem and all points in a rigid plate rotate about identical poles and in the same angles (e.g., Cox and Hart, 1986). However, its points have different poles and/or different rotation angles when a plate undergoes non-rigid deformation. Usually, exact description of motions of points in a real non-rigid plate is too complicated to solve. Dividing a large non-rigid plate into many small rigid plates could be an approximate method to reconstruct their motions. But clearly this method is not exact enough to describe continuous deformation over wide regions. We suggest that rotational pole and angle functions, where that the rotational poles and angles vary continuously with positions of points, are used to approximately describe continuous deformation. Clearly, the formats of the functions depend on the deformation pattern of a deformed plate.

Deformation of the Eurasian plate and the region around the Arctic (except for the marginal basins) can be divided into two parts: (1) "local" thickening of the LTP that results from the diffuse boundary effect of the collision between the Indian and Eurasian plates, and "local" opening of the Eurasian Basin that results from CW rotation of the North American plate relative to the Eurasian plate; (2) "background" diffuse deformation that results from background (or far-field) effects of the collision. The first part can be treated in similar ways to many previous reconstructions (e.g., simply NNE-shortening). The far-field effects that were responsible for rifting of the marginal pull-apart basin system of the NW Pacific will be the focus of this study.

The total diffuse deformation (local and background diffuse deformation together) of the east Eurasian Plate and region around the Arctic (EUAR) can be assumed to have moved in NNE-NE-ENE-EW complex arcuate circuits, progressively approaching North America (Fig. 1), rather than in great or small circles. This is due to resistance from the Pacific oceanic plate that was much smaller than resistance from North American continental plate. In other words, points on the South China block moved NNE; points in the arctic region moved in directions more-or-less parallel to the strikes of transform faults in the Eurasian Basin that was spreading; points in the other regions moved in compatible directions. Magnitudes of motion of points along the line (named as MRLr. Fig. 1) in front of the eastern syntaxis of the collision belt are the largest. The motion magnitudes of other points on the either side of the MRLr decrease both toward the NW and SE. Hence, the region NW of the MRLr is sheared CCW relative to the west Europe, Greenland and northern North America. The region southeast of the MRLr is sheared CW relative to the most eastern margin of the plate. If local thickening in the LTP is subtracted from the total diffuse deformation, the residual diffuse deformation is the background, diffuse deformation.

For simplicity of modeling the background, diffuse deformation, it is assumed that points of the largest movements from any given time to present follow a single curve, which goes from about the northern end of the East Vietnam fault through the west of the North China block, to the region north of Okhotsk Sea (Fig. 1). The curve is here termed the Maximum Rotation Line (MRLb). Points of the MRLb neither have the same rotational pole nor the same rotational radius, as it is actually not a circle on globe. For its southern segment, the average strike is about NS directed and its average pole should be located to the east of the southern segment. For its northern segment, the average strike is east-west and its average pole should be located to the south of the northern segment. If the western Europe to northeastern North America region undergoes little non-rigid deformation, then rotation angles of points in the region NW of the MRLb progressively decrease from maximum values along the MRLb to zero along the west Europe and northeastern North America regions. The region southeast of the MRLb undergoes dextral trans-deformation and rotation angles of the points of this region progressively decrease from the maximum along the MRLb to the proto-arcs of the NW Pacific.

To quantitatively describe this deformation model, many different rotation functions have been attempted. The final model presented here makes simple assumptions that: (1) the finite rotational poles of all the points are located along the equator, and the longitudes of these poles was directly proportional to latitudes of all points considered, and (2) finite rotation angles of points NW of the MRLb were inversely propositional to the arc distances from the poles to the points; finite rotation angles of points of the region southeast of the MRLb is inversely propositional to the arc distances from the poles to the points. If the finite-rotation poles, rotation radius and rotation angles of the points along the MRLb are given, we can use this quantitative model to make a reconstruction of positions of the Eurasian and North American plates through modifying the traditional rigid plate reconstructions for any given time.

### 3.3. Plate reconstructions of development of the marginal basins

Using the non-rigid plate reconstruction method for the Eurasian and N. American plates (see **Appendix**), we present plate reconstructions of the marginal basins of the NW Pacific at four key times: 50, 35, 15 and 5 Ma. All these plate reconstructions will start from rigid plate reconstructions in African hotspot fixed reference frame that are made by Torsvik et al. (2008), and then modify the rigid-plate reconstructions of the Eurasian plate and the region around Arctic using the above quantitative non-rigid deformation model. The reconstruction of the marginal basins is based on the analyses of their rifting history and kinematics that has been stated in Xu et al. (2012) and the approximate reconstruction of the LTP is based on the above analyses of the shortening history.

#### 3.3.1. Plate reconstruction at 50 Ma

Northward motion of the South China block since 50 Ma have been estimated using three independent aspects as stated above: (1) 10.04° to 11.25° relative to Samba Island of the east Java arc, as is indicated by the pull-apart amount of the South China Sea basin and the JMCS basin system, (2) more than 1000 km by the paleomagnetical data, and (3) more than 1200 km by the shortening deficits in the easternmost LTP. Exact motion of this block since then cannot de determined at present level of knowledge. If the rotation angle and rotation radius since 50 Ma along the MRLb (Fig. 1) are assumed to be 11.25° and 70° respectively, the finite rotation angle along the MRLb is about 12°. The pole of points of the maximum latitude (90°, the North Pole) is assumed to have located at (140°E, 0°) and the pole of the points at N10° (latitude) at (180°E, 0). By using these rotational parameters the positions of the EUAR at 50 Ma are reconstructed (Fig. 5).

An important problem with this reconstruction of the EUAR is whether or not the non-rigid deformation of the EUAR is close to plane strain because the Cenozoic topography indicates that no large convergence or extension resulted from this background diffuse deformation (except the "*local*" thickening in the LTP as well as in the Stanovoy Range and the transitional zone between the Eurasian and North American plates). To clarify this problem, contours of ratios of areas of strain ellipses to areas of the unit strain circles are compiled (Fig. 6). Figure 6 shows that all the ratios are close to one, that is, plane strain, though there exists small extensional (ratios less than 1) and compressional (ratios larger than 1) deformation in some regions. The small extensional strain in some regions in Eurasia could have occurred, for example, in the west Siberia Basin to the Canada Basin region. The small extensional deformation in other regions in Eurasia (e.g., in Europe) could not have occurred, but this extension might be compensated by collision of the Arabian plate with east Eurasia. The small extensional deformation along the western North America could not necessarily have occurred, but could have been demolished by the thickening along the Cordillera, which was caused by convergence between the North American and the paleo-Pacific (Farallon and Kula) plates. Dextral transpressional deformation along east Asia and very small uplifts could have occurred. In addition, local small dextral pull-apart basins (e.g., the Bohai Gulf basin) could have overprinted a slightly uplifting eastern Asia.

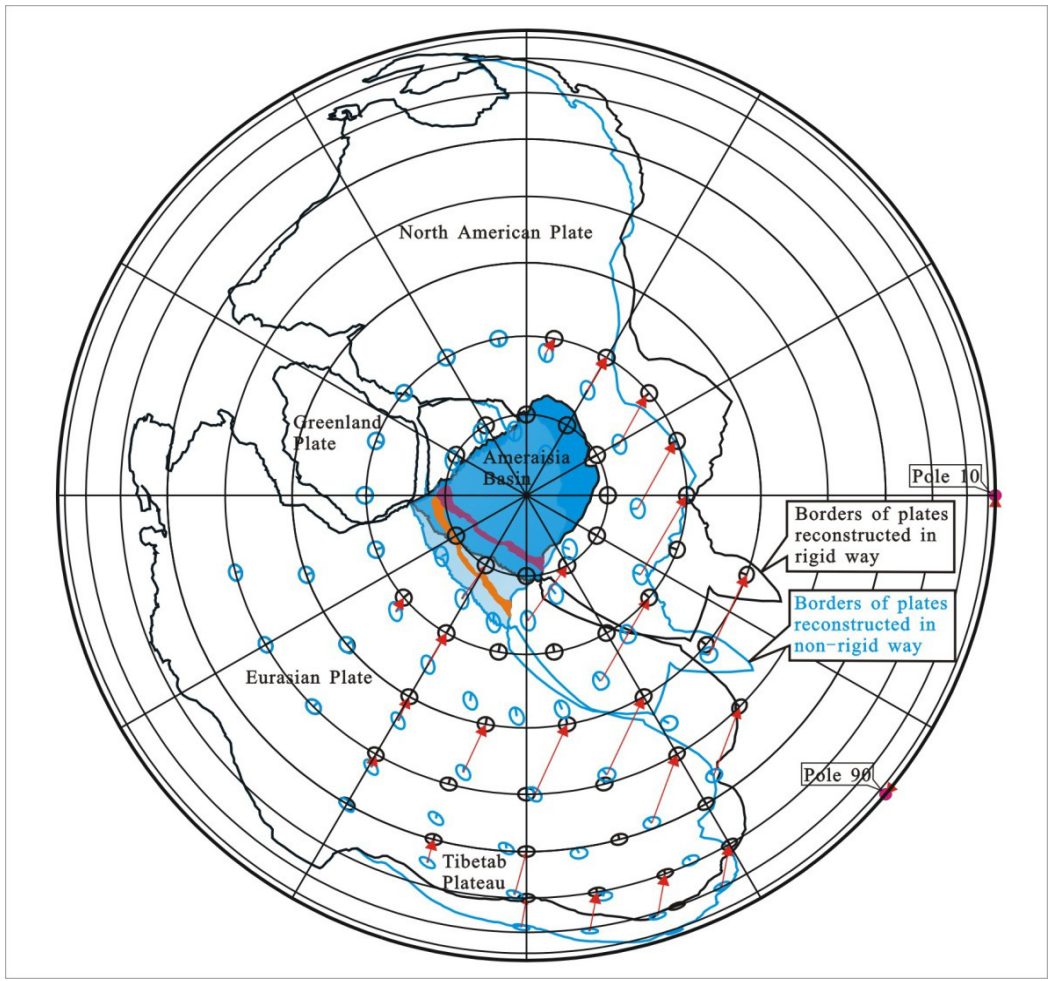

Fig. 5. Reconstruction of the defuse background deformation of the EURA at 50 Ma (see text for details). Black lines represent the positions of plates in rigid plate reconstruction in African Hotspot Fixed Reference Frame (Torsvik et al., 2008). Blue lines are positions of the plates in the reconstruction by this paper. Pole 10 and Pole 90 denote the Euler poles of rotation of points located on 10 degree and 90 degree latitudes on the rigid Eurasia and North America plates for the background defuse deformation reconstruction, respectively. Black small ellipses are unit circles (one degree arc distance radius) on the rigid Eurasia plate and blue ellipses are restored from the unit circles in the non-rigid way. Note that the rigid plate reconstruction in African Hotspot Fixed Reference Frame doesn't show "local" contractional deformation of the Tibetan Plateau and its adjacent regions.

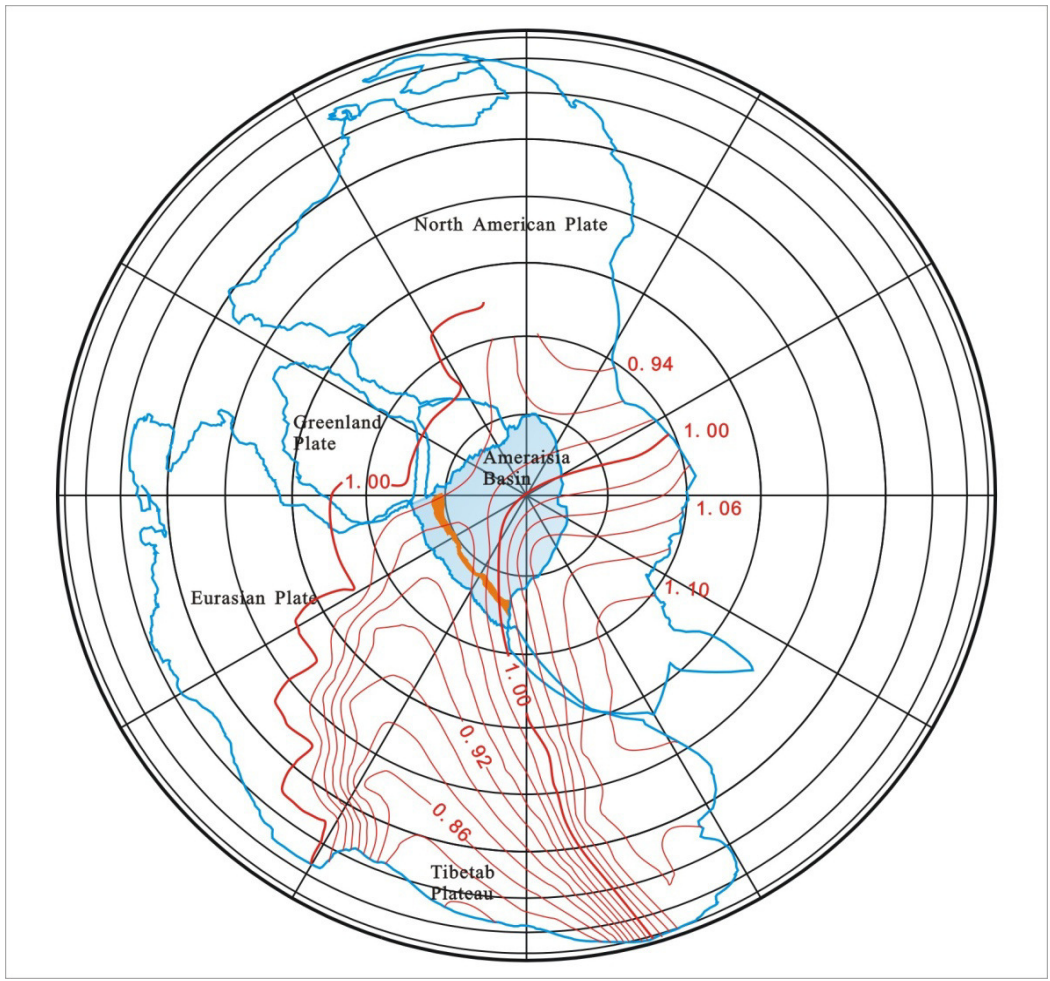

Fig. 6. Contours of area changes in the EURA for the background non-rigid deformation since about 50 Ma (see text for details). The values labeled on the contours (red lines) are ratios of the restored ellipses' areas to the areas of the circles as stated in Fig. 4. The contours clearly show that the non-rigid deformation are close to the plane strain. The regions delimited by the less than one-labeled contours incur slight transtensional deformation and those with larger than one-labeled contours undergo slight transpressional deformation. Light green lines were coastlines of the Eurasia to Alaska. See Fig. 4 for other symbols

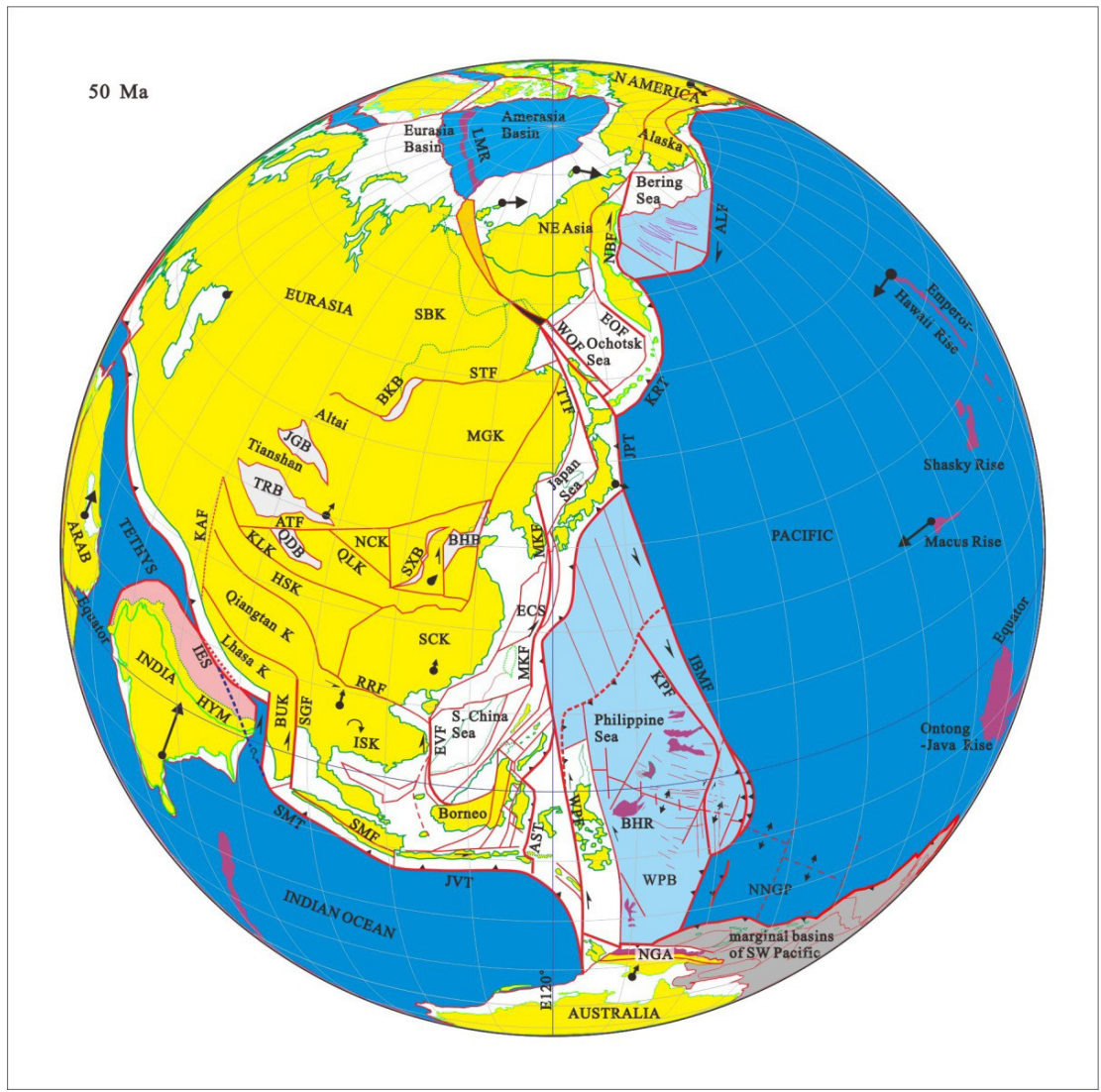

Fig. 7. Plate reconstruction of the NW Pacific margin and the Large Tibetan Plateau at 50 Ma. Pink area north of India block is north part of the Great India that has been subducted under the Tibetan Plateau and accreted to Himalayas (HYM). Black arrows represent motion-vectors of starting points from 50 to 35 Ma. Dark blue dashed line could be an alternative segment of the boundary between Tethys and Asia. Red dotted lines in oceanic crust of marginal basins are magnetic lineations. Deep yellow and dark blue areas denote overlap and gap between the Eurasian and North American plates, respectively. Note that the local deformation in Stanovoy and the transitional zone between Eurasian and North American plates are not reconstructed (see Part One for detail). Possible small displacements of the South China Mongolia and North China blocks relative to the Siberia block are not reconstructed, except the background deformation. Marginal basins of SW Pacific (dark gray) are not reconstructed. See text for details. Abbreviations (alphabetically): AST= ancient East Sunda Trench, ALF= Aleutian fault, BHR=Benham Rise (possibly formed at 49.4 Ma), IBMF= Izu-Bonin-Mariana fault, KPF=Kyushu-Palau fault, NGA=New Guinea Arc, NNGP = North New Guinea plate. All other abbreviations and legends are same as these in Figure 1.

The Philippine Sea plate at 50 Ma was located about 20° south of its present position as indicated by paleomagnetic data (mentioned in Xu et al. (2012)). It is assumed that the finite rotation pole and finite rotation angle of the WPB at 50 Ma are at (210° E, 12 ° N) and 20°, respectively.

Taking the above kinematical analyses into account, the plate reconstruction of the marginal basins of the NW Pacific and the LTP at 50 Ma is presented (**Fig. 7**). Some related key issues are summarized as follows.

**(1) Before 50 Ma.**

In the NW Pacific, there were three oceanic plates: North New Guinea, Pacific and Kula plates (e.g., Hilde et al., 1977; Ben-Avraham, 1978; Seno and Maruyama, 1984; Lewis et al., 2002). There was a NW-trending spreading axis between the Pacific and North New Guinea plates, across which NNE-striking transform faults developed. To the north, these transform faults linked to NNW-striking transform faults within the western Pacific Plate.

Along the continental margin of the NW Pacific, there had possibly already existed right-lateral, stepped mega-fault zones that include the East Vietnam, East Korea-Manila, Tartar-Tanakura and East Okhotsk faults because they had controlled distribution of the late Mesozoic strata (Xu et al., 2012). The Vietnam fault zone was short and did not cut through the Sunda Shelf possibly because the west Borneo-East Malay block docked on the proto-Indochina-South China block in the late Mesozoic (e.g., Audley-Charles et al., 1988) and the southern half of the long proto-East Vietnam fault zone was overlaid with this block. The narrow proto-South China Sea including the Sibu-Miri and Sabah basins in northern Borneo survived.

The geometry and positions of the basins and blocks in the Banda Sea region are very difficult to determine, because this region has been intensively deformed, and because its geology and geophysics are still poorly known although several evolutionary scenarios have been put forward (e.g., Charlton, 1986; Lee and McCabe, 1+986; Hall, 2002; Hinschberger et al., 2005). Many papers suggest that the micro-continental blocks including the east and south arms of Sulawesi, the Banda Ridge, Buru-Seram and Sula were parts of the Australian continental block (e.g., Lee and Lawver, 1995; Hall, 2002). However, their affinities to and times of rifting away from Gondwanaland are uncertain (e.g., Hartono, 1990; Ali et al., 1996; Vroon et al., 1996) and their paleomagnetic data are contradictory (e..g., Johnsma and Barber, 1980; Aili and Hall, 1996). The Banda Sea region might be a mosaic of continental crusts and the Mesozoic and Cenozoic oceanic crust because some ophiolites (e.g., Buton and Seram ophiolites) are Mesozoic (Hall, 2002). "Coincidental" alignment of the Java arc with the Inner Banda arc at present has long been probably hereditary since the proto-Inner Banda arc that is as old as the proto-Java arc formed. There possibly was proto-New Guinea-Inner Banda-Java arc between the Indian and Pacific oceans and SE Asia, which linked with the Sumatra arc (Fig. 7). The East Sundaland trench, a plate boundary, was probably located between Sundaland and the Banda Sea region (rather than Indian Ocean).

There are two possibilities regarding the position of the south margin of the Lhasa block (Fig. 7): it trended in north-NW, roughly parallel to the Sumatra Trench, this means that there is no "coincident" large gulf in the south margin of the Eurasian Plate to "wait for" the coming India block "to fill"; there is a "coincident" large gulf west of the Burma Block, which might have formed by collision of the Lhasa and Qiangtan blocks with Asia during Mesozoic (or other causes). Considering the possible initial collision time and shortening deficits as mentioned above, the second possibility is adopted.

The Tibetan Plateau region possibly had no high average altitude as suggested by Dewey et al. (1988). The continental arc along the south margin of the Eurasian plate might be thin like the Sumatra and Java arcs at present. Although crust of the Lhasa and Qiangtan blocks could have been thickened in the Mesozoic (e.g., Murphy et al., 1997), crust of the Qiangtan block might have returned to about normal thickness just before the collision between India and Asia because of erosion. As mentioned above (see Fig. 3 for details), if the LTP had had high altitude or significantly thickened crust just before the initial collision, there would have been unrealistic shortening deficits in it. The Lhasa block could be as high as the present Sumatra arc.

The overlap and gap between the Eurasian and North American plates in NE Asia is similar to these in the rigid plate reconstruction. This indicates the diffuse deformation around the boundary between the two plates. Some special rotational function for the points in the eastern and northeastern Asia could be found to eliminate the overlap and gap, but this is difficult to determine and will not significantly help to understanding of plate reconstruction of the marginal basins of the NW Pacific.

**(2) At 50 Ma and between 50 and 35 Ma**

At about 50 Ma, the Australian continental block collided with the New Guinea arc (e.g., Hall, 2002), the polarity of the subduction zone along the New Guinea arc inverted and the New Guinea Central orogen began to form. This collision was in about NS direction and formed a sinistral transpressional "pop-up" in the Pacific and N New Guinea plates, entrapping the spreading axis between the two plates and NE- and NNW-striking transform faults. The pop-up structure was the proto-Philippine Sea plate (Fig. 7). According to paleomagnetic data (Xu et al, 2012), the Philippine Sea plate (PHP) was located about 20° south of its present position. The PHP was surrounded by the sinistral transpressional fault zones: the Negro-Cotabas trans-thrust（the west Philippines fault, WPF）west of Philippines-Halmahera in the west, the North Philippine Sea trans-thrust within the Pacific plate in the north, Izu-Bonin-Mariana trans-thrust (IBMF) (possibly a pair of fault zones along the Mariana segment) in the east, Yap thrust (YPF) and Palau thrust (PLF) in the south. The Philippine Trench fault (PTF) was a sinistral transpressional one that was parallel with the WPF. Within the PHP, linkage of two large NNW-striking transform fault with one large NNE-striking transform fault (a segment of the Sofugan fault) formed the Kyushu-Palau Fault (KPF). The KPF and the Sofugan fault divided the PHP into three sub-basins: West Philippine Basin (WPB), the proto-Parece Vela Basin (PKB) and proto-Shikoku Basin (SKB). The SKB was very narrow and would be completely occupied by later Izu-Bonin Ridge and Kyushu-Palau Ridge. **With** continuation of compression of the northward-moving Australia block and New Guinea arc, the PHP gradually formed into a more mature transpressional "pop-up" structure. The Pacific plate subducted below the PHP along the IBMF-YPF-PLF and the Izu-Bonin-Mariana-Yap-Palau arcs (IBMR-YPR-PLR) began to form. The IBM subduction zone has a steep dip angle at present possibly because it resulted from a transpressional thrust rather than the subducting Pacific plate was older and denser. Igneous rock extruded and intruded along the zig-zag KPF, and Kyushu-Palau Ridge (KPR) began to form. The KPR could include some island-arc geochemical characteristics because of being close to the subduction zone. Similarly, the Gagua Ridge began to form along the northern segment of the PTF. The NW- and NS- striking conjugate faults developed within the PHP, and basaltic magama intruded and extruded along some of these faults. The basaltic rocks along the faults could form the linear sea-floor fabrics and magnetic anomalies. These magnetic anomalies of different orientations did not indicate seafloor spreading in different directions. Up to near 45 Ma, the West Philippine Sea could have fully opened, because the oldest age of basalt samples from Benham Rise which extends in NE-trending direction close to the spreading axis (central Philippine Sea fault, CPF) is 49.4 Ma (see summary of Deschamps and Lallemand, 2002). However, if the West Philippine Sea Basin (WPB) had continued to spread to 33 Ma as some authors suggest (e.g., Hilde and Lee, 1984), this active spreading must have happened within the pop-up structure and the oceanic crust subducted into the Philippine Trench. From 50 to 35 Ma, the eastern ridges of the PHP continued to undergo local CW rotation because the northward motion of the PHP is much larger than the northward component of the motion of the Pacific plate and dextral shear along the IBMR, YPR and PLR took place. In the meantime, the northern segment of the IBMT retreated towards the Pacific in small scale and the spreading behind the IBMR began.

At about 50 Ma, the Indian continent began to collide with the Eurasian plate and impinged east Asia to move in complex, NNE-NE-ENE-EW arcute circuits (stated above). This happened at the same time as major marginal basins began to open as dextral transtensional basins. The Aleutian and Komandorsky sub-basins of the Bering Sea basin was trapped as a dextral pull-apart in the Pacific plate (or Kula plate) with the Aleutian Trench fault acting as its eastern and southern boundaries. But the Japan Sea and Okhotsk Sea basins underwent much weaker rifting than the northern margin of the South China Sea because direction of the Amurian block moved at a high angle to the main boundary faults of the former two basins. An exception is that Celebes Sea might result mainly from the active spreading although the dextral transtension might trigger start of its spreading.

Geometry of the Indochina-Sumatra block (ISK) and the Borneo-Java Arc region at 50 Ma is reconstructed through 25 degrees of CCW rotation of the present ISK in the way as suggested in the companion paper (Xu et al., 2012). From 50 to 35 Ma, the ISK continued to rotate CW with the Indian plate pushing against Asia. The south shelf of the Andaman Sea basin, west of the ISK, began to rift as a dextral pull-apart margin.

The Lhasa block as a continental arc probably had higher elevations than other blocks at 50 Ma. The LTP began to uplift since 50 Ma. The uplift was mainly attributed to contraction of crust of the Tibet Plateau itself rather than the underplating of the Indian continent. But the North China block (the Alxa region) might be subducted below the Qilianshan block and the Qilianshan block. By 35 Ma, the Lhasa and Qiangtan blocks had nearly reached its present altitude and the northern Tibetan Plateau might have significantly uplifted.

The India-Asia collision might have triggered the Eurasian Basin's spreading along the Garkel Ridge if the initial collision time were 55 Ma. However, the initial collision time is possibly 50 Ma and it is difficult to correlate the collision with the beginning of the Eurasian Basin's spreading (at about 56 Ma). As mentioned above, 55 Ma of the initial collision time cannot totally be excluded and the correlation between the collision and the rifting start of the Eurasian Basin could be an unsettled problem. Interestingly, the Gakkel Ridge was probably aligned about parallel with the North Atlantic Ridge at around 50 Ma. The Mesozoic Amerasia basin underwent strike-slip deformation since 56 Ma (e.g., Rowley and Lottes, 1988) and its complex magnetic anomalies might have been attributed to the reworking of the Cenozoic non-rigid, trans-deformation.

### 3.3.2. Plate reconstruction at 35 Ma

The spreading time of oceanic crust of the South China Sea is problematical, as discussed in Xu et al. (2012). Reconstruction here adopts a "conservative" alternative, that is, its spreading time is about 32 to 15 Ma (Briais et al., 1993). The JMCS transtensional system also fulfilled its rifting during this period. This means that the South China block moved northward by nearly 900 km along the central part of the South China Sea and the JMCS system. Therefore, it is assumed that the finite rotation angle along the MRLb is 10° and the rotation radius along the MRLb (Fig. 1) is assumed to be 70°. Similar to reconstruction at 50 Ma, the pole of points of the maximum latitude (N90°, the North Pole) is also assumed to have located at (140°E, 0°) and the pole of the points at N10° (latitude) at (180°E, 0°). The PHP is assumed to have moved northward about in step with the northward-moving Australia Block from 50 to 35 Ma and the finite rotation pole and finite rotation angle of the WPB for 35 Ma are at (210°E, 12°N) and 16°, respectively.

Plate reconstruction of the marginal basins of the NW Pacific and the LTP at 35 Ma is shown in Fig. 8. The main evolutionary characteristics from 35 to 15 Ma are summarized as follows.

The Pacific plate kept moving westward. Spreading occurred in the Caroline Sea west of Pacific from at least 36 to 28 Ma (Hegarty et al., 1983; Hegarty and Weissel, 1988). The Caroline Ridge formed during the same time as this spreading. The Caroline Ridge and sea constituted the Caroline plate. Where was this plate and what relationship between it and Pacific plate is a puzzling problem. Many authors believe the Caroline Sea basin was one of the back-arc basins of the NW Pacific and had been located south or southeast of the PHP since its formation (e.g., Hall, 2002; Gaina and Muller, 2007). There would have been a large convergence (more than 2000 km) between the Pacific and Caroline plates if these authors' ideas are correct. However, the southern segment of the boundary between the two plates is the Lyra Trough, which is now an inactive graben, is very small and there is not any evidence (e.g., ancient arc) to indicate existence of a large convergence along this trough. The northern segment of the boundary cannot be found in topography. The Mussau Trench could be an alternative boundary between the two plates, but this trench is also small, and convergence along it is less than 100 km and occurred during late 1 Ma (Hegarty et al., 1983). So the reconstructed Caroline Ridge (plate) moved in step with the Pacific plate. The Caroline Sea basin might be a major oceanic plate most of which has subducted below the PHP and the Australia-SW pacific marginal basin region.

The PHP continued to move northward a little slower than the Australia continental block and limited convergence between the PHP and the Australia continental block occurred. The PHP further evolved into a mature sinistral pop-up structure. Meanwhile, the great eastward movement of the northern IBMR together with the eastward-moving Japan arc caused fan-like spreading behind the ridge since 35 Ma. At about 30 Ma, the spreading entered the proto-Shikoku Basin and the present Shikoku Basin began to open. The spreading propagated southwards and reworked the proto-Parece Vela basin along about NS-striking faults and the Parece Vela Basin became wider. According to the pattern of magnetic lineations, the spreading occurred in east-west direction from 35 (or 30) to 19 Ma and in NE direction from 19 to 15 Ma (C6 to C5c) in both the Shikoku and Parece Vela basins. Change in direction of the spreading possibly indicated that the interaction between related plates changed. It is postulated that there was more intensive convergence between the PHP and the Australian continental block, and sinistral trans-deformation occurred within the sinistral pop-up structure of the PHP from 19 to 15 Ma. Superposition of the sinistral trans-deformation on the east-west-spreading resulted in sinistral transtensional spreading in the two sub-basins. The PHP as a whole moved northward relative to the Pacific plate and the local dextral strike-slip deformation along its eastern ridges continued from 35 to 15 Ma.

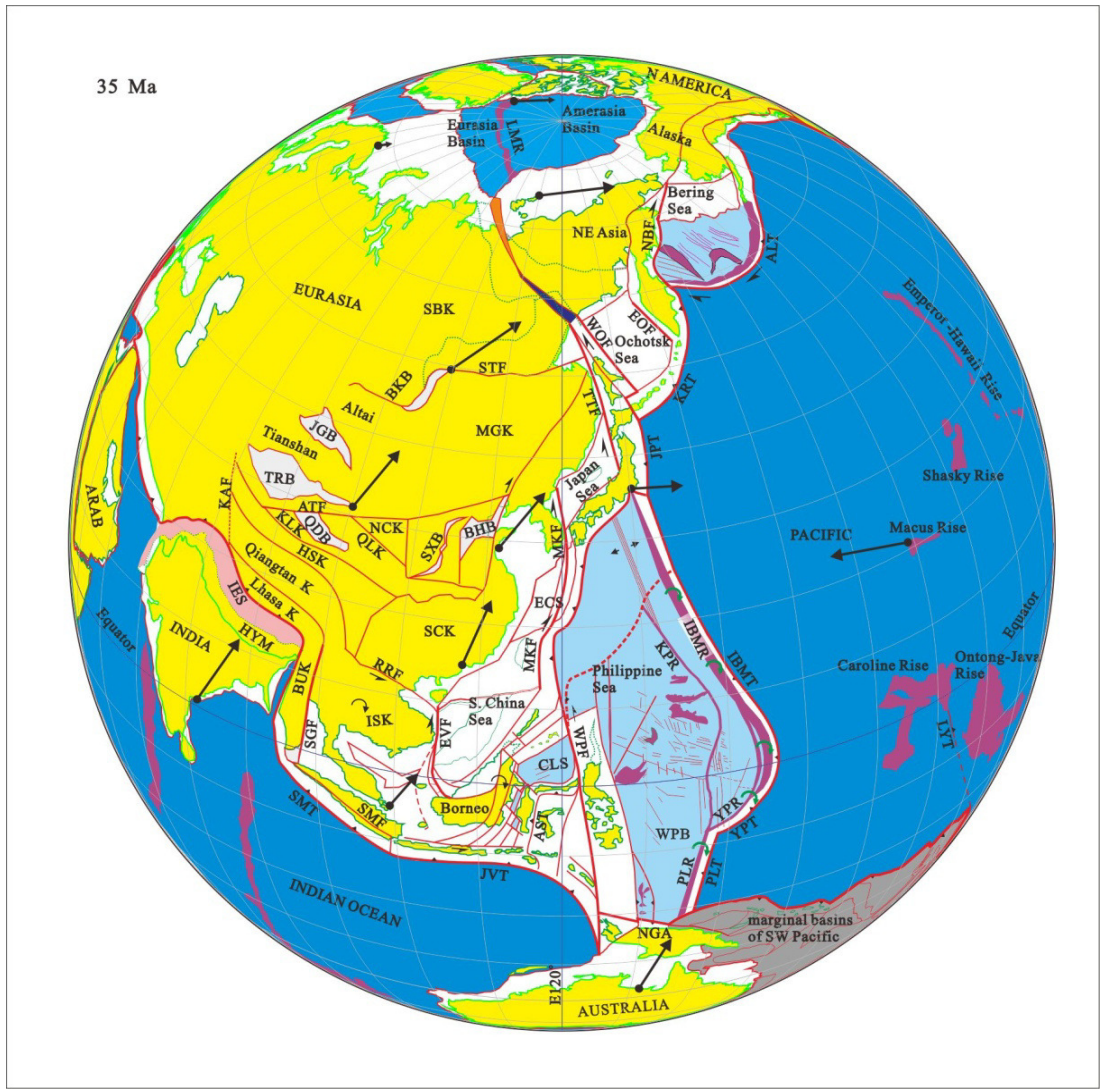

Fig. 8. Plate reconstruction of the NW Pacific margin and the Large Tibetan Plateau at 35 Ma. Balck arrows represent motion-vectors of starting points from 35 to 15 Ma. Green circular lines with arrows on the IBMR, YPR and PLR indicate the local clockwise rotations. Note that the local deformation in Stanovoy and the transitional zone between Eurasian and North American plates are not reconstructed and the Kuril Trench "retreated" southeast-ward from 35 to 15 Ma (see Part One 3.7.3 for details). See text for details. Abbreviations (alphabetically): AST=ancient East Sunda Trench, ALF= Aleutian fault, IBMT= Izu-Bonin-Mariana Trench, LYT=Lyra Trench (Trough), PLT=Palau Trench, YPT=Yapu Trench. All other legends and abbreviations are same as in Figures 1 and Figure 7.

With large NNE-NE-ENE-EW movement of east Eurasia and region around the Arctic in steps progressively approaching North America, the marginal basin system of the NW Pacific entered intensive rifting period from 35 to 15 Ma. However, differences in rifting intensity or pull-apart amount of these basins existed because of different local plate tectonic settings (Fig. 8). At about 32 Ma, seafloor spreading began in the South China Sea basin and possibly in Japan and Okhotsk seas basins as well as in the JMCS basin system. The northern margin of the South China Sea and the Shelf Basin of the East China Sea generally entered the weaker rifting stage some like passive continental margins when the seafloor spreading proceeded, but intensive rifting continued in the western continental margin of the South China Sea in order to keep kinematic balance with the seafloor spreading in the Central and Southwest sub-basins. The Kormandorsky Basin in the Bering Sea spread for intensive dextral shear along the southwest segment of the Aleutian Trench and sub-basins in the outer Bering Shelf underwent dextral transtension.

Geometry of the ISK and the Borneo-Java Arc region at 35 Ma are reconstructed through 20° of CCW rotation of the ISK. From 35 to 15 Ma, the ISK continued to rotate CW with the Indian plate pushing on the Eurasian plate . In the Andaman Sea region, dextral transform margin-type rifting was active in the Mergui basin with principal fault being the Sumatran Fault system, and both the transform margin-type rifting and the dextral pull-apart rifting were coevally active in the Andaman Sea Basin when the Sumatra fault rotated CCW enough and obliquity of subduction of the Indian plate motion along the Sumatra trench was enough to trigger the dextral displacement to take place on the Sumatra fault system and the Mottawi fault. The rifting in the central Andaman Sea basin could be unusual seafloor spreading that formed Alcock and Sewell oceanic plateaus from 35 to 15 Ma.

Shortening and uplifting rates in the LTP were much lower during this period than the last period. Elevation of the LTP generally remained constant. Only the Himalayas and region around Tarim basin could uplift.

Very wide-spread, latest Oligocene unconformity around 25 Ma in NW Pacific margin and the LTP might be attributed to an unknown plate re-organization. This plate re-organization could result in the jump of the spreading axis of the South China Sea.

### 3.3.3. Plate reconstruction at 15 Ma

At 15 Ma, the eastern Eurasian plate ceased to move NNE- to ENE. The Eurasian plate and the region around the Arctic is reconstructed as in the rigid reconstruction (Torsvik et al, 2008). The PHP is assumed to have moved northward a little slower than the Australia continental block from 35 to 15 Ma because a small amount of convergence between the PHP and the continental block occurred during this period. Its finite rotation pole and finite rotation angle for 15 Ma are at (210° E, 12° N) and 9°, respectively. The plate reconstruction of the marginal basins of the NW Pacific and the LTP at 15 Ma is shown in Figure 9. The main evolutionary characteristics from 15 to 5 Ma are summarized below.

Between 15 and 5 Ma, the PHP continued to move northward. The seafloor-spreading in SKB and PVB also ceased because the Japan arc together with the IBMR did not significantly migrate eastward. The WPB was reworked and the magmatic activity (e.g., 10 Ma basalt along the CBF) occurred in this period. The Caroline Ridge was close to the IBMR.

All the major basins basically stopped rifting and began to enter the post–rifting period from 15 to 5 Ma. Most of them were still be reworked by tectono-magmatism and the basaltic flows often occurred forming the seamounts. Sinistral transpressional inversions developed.

Geometry of the ISK reached its present geometry at 15 Ma because the South China Sea no longer stretched in north-south-direction. The Borneo-Java Arc region at 15 Ma could have been reconstructed to its present geometry, but a presumed 5 degrees of CW simple shear along north-south-trending vertical planes is added to the its present geometry because of its CCW rotation and convergence with the Dangerous Ground-Reed Bank (Nansha) block in the South China Sea (after 15 Ma). In the Andaman Sea region, north-NW-directed weak transtensional rifting on the Alcock and Sewell plateaus and NW-directed weak transform margin type rifting continued in the Mergui Basin from 15 to 5 Ma.

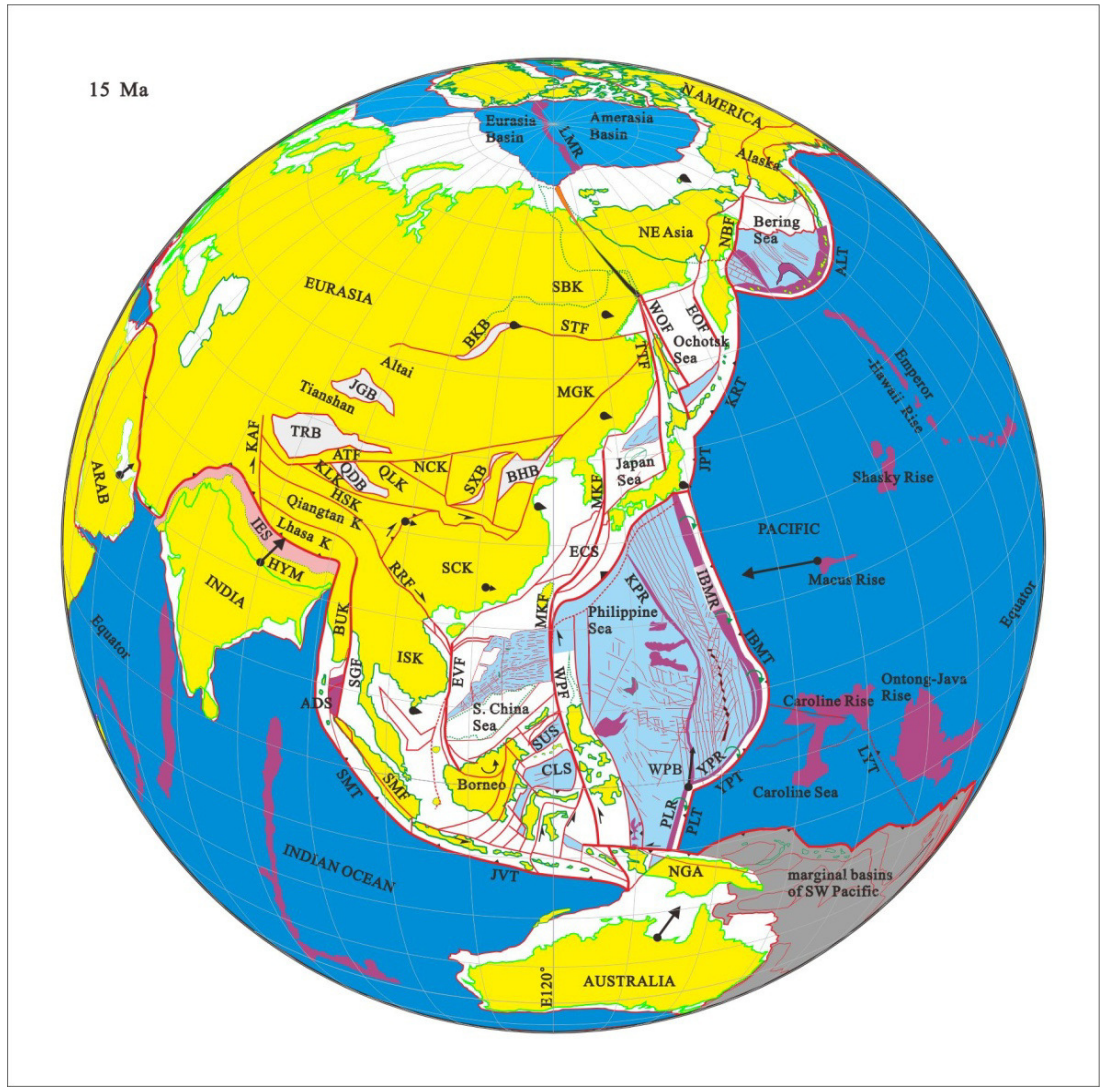

Fig. 9. Plate reconstruction of the NW Pacific margin and the Large Tibetan Plateau at 15 Ma. Black arrows represent motion-vectors of starting points from 15 to 5 Ma. See text for details. All legends and abbreviations are same as in Figures 1 and Figure 7.

The more rapid shortening and uplift in the LTP resumed since 15 Ma because the Eurasian plate became basically stable. From the Indian Ocean northward to the region around Tianshan, the intensive compressional deformation rejuvenated. The Himalayas rapidly uplifted. The Lhasa and Qiangtan blocks, however, might not have exhibited large-scale uplift and incurred no significant shortening. The Hoh Xil-Sangpan-Ganze block and the Altyn and Tianshan regions also began to uplift rapidly. The eastern Tibetan Plateau commenced to thrust over the western South China block along the Longmenshan belt. For the rapid convergence and uplift, most of the NS-trending grabens formed on the Tibetan Plateau, as is a typical indicator of the rapid shortening and uplift. The Karakorum fault rejuvenated as a right-slip fault.

A plate dynamic problem is why the east Eurasian plate ceased to move NNE to ENE since around 15 Ma while the Indian plate was still colliding with it. Rates of the Indian plate relative to the "fixed" Eurasia and the African hotspot fixed frame did not evidently change (Fig. 2). It is speculated here that there were two causes responsible for this: (1) the main part of North American plate (except the region of the Arctic) had rotated to the eastern front of the eastward-moving EUAR and resisted the EUAR's further eastward-moving; and (2) the convective state of the upper mantle below the Eurasian and North American plates began to change and this might have reduced the northward movement of both the Indian and the eastern Eurasian plates. The second cause might have played more important role in this change.

### 3.3.4. Plate reconstruction at 5 Ma

At 5 Ma, the Eurasian plate and the region around the Arctic is also reconstructed in the same way as in the traditional reconstruction (Torsvik et al, 2008). The PHP at 5 Ma is reconstructed with its finite rotation pole and finite rotation angle being at (155.53°E, 32.73°N) and 7.54°, respectively. Plate reconstruction of the marginal basins of the NW Pacific and the LTP at 5 Ma is shown in Figure 10. The main evolutionary characteristics from 5 to 0 Ma are summarized as follows.

The Caroline Ridge collided with the southern segment of the IBMR and the Mariana Basin began to open. The West Mariana Ridge continued to move relative to the IBMR from 5 Ma to present and seafloor spreading in the Mariana Basin is occurring today. The PHP generally moved NW-ward. The northern Philippine arc collided with the South China continent and the Taiwan orogen commenced its formation. The northern Philippine Arc pushed the Taiwan-Sinzi rise to move north-south- to north-NW-ward relative to the Ryukyu Arc during Pliocene and Quaternary. This could cause the local dextral transtension to have occurred in the Okinawa Trough.

All the marginal sea basins basically stopped rifting from 5 to 0 Ma, but more intensive, though very weak, rifting episodically occurred than the last period (15 to 5 Ma).

The ISK kept the same geometry as in the last period. The Borneo-Java Arc region rotated CCW, which continued its further convergence with the Nansha block since 5 Ma. Much more intensive north-NW-directed transtensional rifting or seafloor spreading occurred in the Central Andaman Basin since about 5 Ma.

The rapid shortening and uplift in the LTP continued following the last period. The Himalayas continued to grow up due to the shortening. The Lhasa block might have uplifted due to the underplating of the Indian plate. The Hoh Xil- Sangpan-Ganze block and the Altyn and Tianshan regions also kept rapidly uplifting. The eastern Tibetan Plateau seemed to have rapidly extruded easterly in this period relative to the last period, as is indicated by the sinistral strike-slip motion along the Qinling and Jicheng faults that intensified rifting in the Shanxi Basin. The NS-trending grabens on the Tibetan Plateau rifted intensively and the right-slip Karakorum fault was active during this period.

A problem is why the weak rifting was more extensive and intensive in this period than in the last period (e.g., rifting in the Shanxi and Central Andaman basins, and more rapid subsidence in some marginal basins and their subbaisins in this period). We speculate that the convective state of asthenosphere below the Eurasian and North American plates once again evidently changed and more rapid northward motion of the Indian plate resumed, if Copley et al. (2010) is right (Fig. 2).

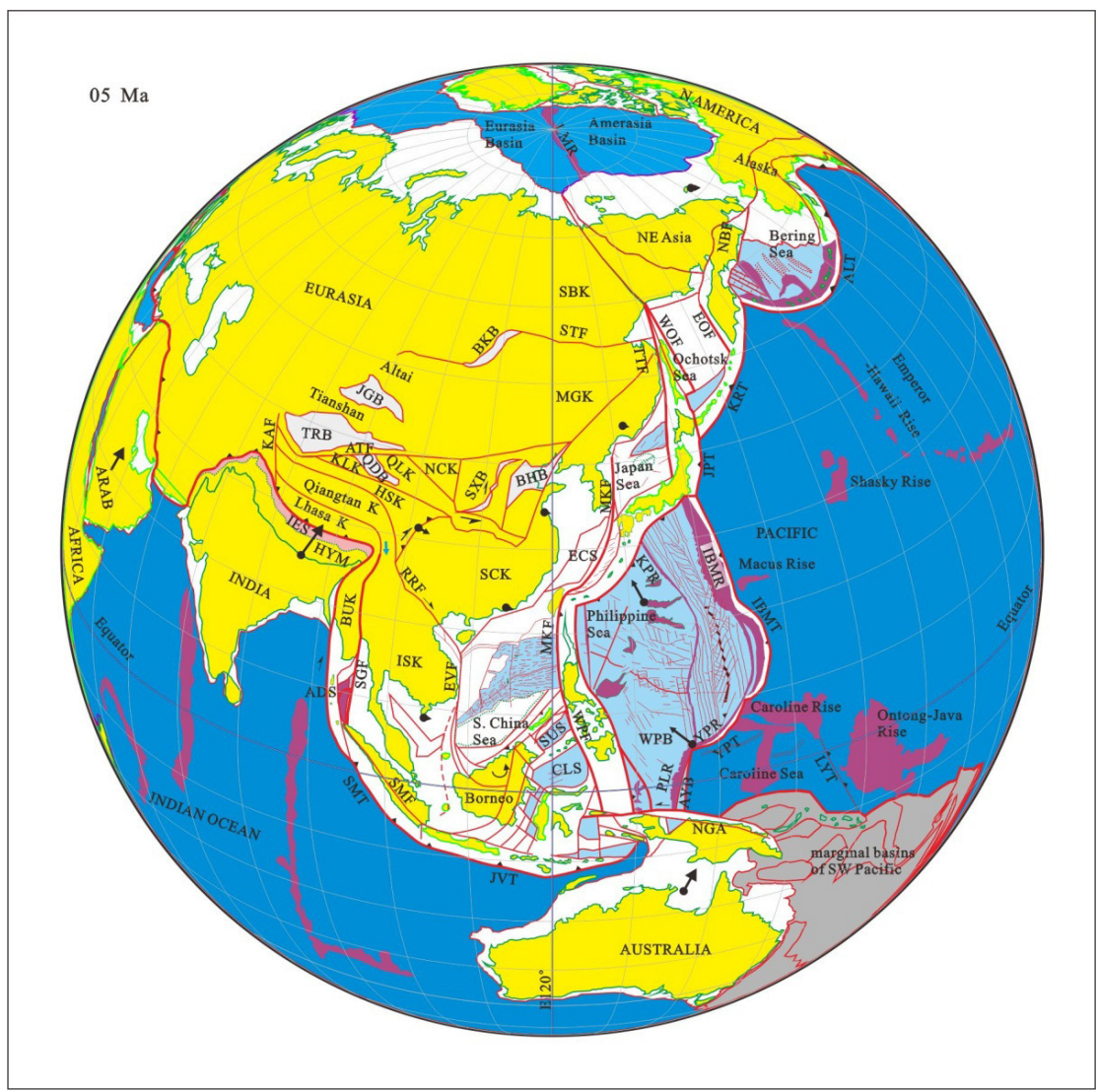

Fig. 10. Plate reconstruction of the NW Pacific margin and the Large Tibetan Plateau at 5 Ma. Black arrows represent motion-vectors of starting points from 5 to 0 Ma. Blue arrow denotes local material flow direction from the Tibetan Plateau to the most northwestern Indochina-Sumatra block. See text for details. All legends and abbreviations are same as in Figures 1 and Figure 7.

## 4. Conclusions

Through plate reconstructions at 50, 35, 15 and 5 Ma of the marginal basins of the NW Pacific and the Large Tibetan Plateau (LTP), the main conclusions are reached as follows.

1. Two lines of direct evidence that include pull-apart amount of the South China Sea basin and JMCS basin system and the shortening deficits in the LTP, and one indirect evidence from paleomagnetic data indicate that there was more than 1000 km of absolute motion of the South

China block and other blocks in east Eurasia and region around the Arctic.

2. The initial collision of the Indian plate with the Eurasian plate was about 50 Ma ago, was asynchronous along the collision belt with full collision occurring between the two continents at about 45 Ma. The uplift history of the LTP since the initial collision can be divided into two first-order stages: (1) the slower uplift stage from 50 to 15 Ma, which contains both rapid and extremely slow uplift sub-stages, and (2) the more rapid uplift stage from 15 Ma to present.

3. The Eurasian plate and the region around the Arctic underwent large, horizontal, nearly simple-shear, diffuse background deformation with the maximum translation of around 1200 km since the initial collision of the India with Asia and this background deformation caused the pull-apart rifting of the marginal basins along the NW Pacific.

4. Rifting history of the marginal basins is closely correlated with the uplift history of the LTP. The two first-order uplift stages of the LTP correspond to the first-order rift and post-rift stages of the marginal basins, respectively.

5. The Philippine Sea basin was trapped as a sinistral transpressional pop-up structure across the Pacific and North New Guinea plates between a NW-trending spreading axis at a position that was 20° south of its present position. Then, it episodically moved northward in a large scale. While the Japan arc migrated eastward in a large scale, the seafloor spreading occurred in Shikoku and the Parece Vela basins. The collision of the Caroline Ridge with the West Mariana Ridge resulted in the opening of the Mariana Basin.